\documentclass[preprint]{aastex}

\usepackage{pstricks}
\usepackage{graphicx}
\usepackage{txfonts}
\usepackage{natbib}
\begin{document}
\title{Transport of magnetic flux from the canopy to the internetwork }

\author{A. Pietarila}
\affil{National Solar Observatory, 950 N. Cherry Avenue, Tucson, AZ 85719, USA }
\and

\author{R. H. Cameron, S. Danilovic and S. K. Solanki\altaffilmark{1}}
\affil{ Max-Planck-Institute for Solar System Research, Max-Planck-Strasse 2, 37191 Katlenburg-Lindau, Germany}
     
\altaffiltext{1}{School of Space Research, Kyung Hee University, Yongin, Gyeonggi, 446-701, Korea}

\begin{abstract}
  Recent observations have revealed that 8\% of linear polarization patches in the internetwork quiet Sun are fully embedded in downflows. These are not easily explained with the typical scenarios for the source of internetwork fields which rely on flux emergence from below. We explore using radiative MHD simulations a scenario where magnetic flux is transported from the magnetic canopy overlying the internetwork into the photosphere by means of downward plumes associated with convective overshoot. We find that if a canopy-like magnetic field is present in the simulation, the transport of flux from the canopy is an important process for seeding the photospheric layers of the internetwork with magnetic field. We propose that this mechanism is relevant for the Sun as well, and it could naturally explain the observed internetwork linear polarization patches entirely embedded in downflows.
       
\end{abstract}

\section{Introduction}

 High spatial resolution and high signal-to-noise spectropolarimetric
 observations, such as those from the spectropolarimeter (SP, \citealt*{HinSP2001}) on-board the
 Hinode satellite \citep{Hinode,Tsuneta+others2008,Suematsu+others2008,Shimizu+others2008} and the full
 Stokes magnetograph IMaX \citep{MartinezPillet+others2010} on-board
 the balloon-borne telescope Sunrise \citep{Barthol2010,
   Solanki+others2010,Berkefeld+others2010, Gandorfer+others2010}, have made it possible to characterize in
 greater detail the photospheric magnetic fields in the internetwork
 (INW) regions of the Sun. For a recent review on small-scale magnetic
 fields see \citet{deWijn+others2009}.

It is generally proposed that the INW fields come from below (e.g., \citealt{WedemeyerBohm+others2009}),
emerging from the high plasma-$\beta$ ($8 \pi \rho / B^{2}$) region where the magnetic
fields can be inductively amplified, into the
higher layers where the plasma-$\beta$ is low.
Possibilities include a source associated with the global dynamo (recycling of flux from decaying
active regions and inductive amplification, cf. \citealt{Ploner+others2001,Hagenaar+others2003,TrujilloBueno+others2004}) or local dynamo action \citep{Petrovay+Szakaly1993}.

 What all these mechanisms above have in common is that the flux
 originates mainly from below the surface. Therefore, one would expect
 the appearance of INW magnetic fields to be associated with flux
 emergence. This has been observed in several instances, e.g.,
 \citet{Centeno+others2007,MartinezGonzalez+others2007,MartinezGonzalez+BellotRubio2009}. Recent
 observations from the IMaX instrument on-board the Sunrise
 balloon-borne telescope allowed \cite{Danilovic+others2010} to study
 in detail linear polarization patches in the INW. IMaX observed the
 full Stokes vector of the photospheric 525.0 nm \ion{Fe}{1} line at
 high signal-to-noise ratio and high spatial resolution. Of the nearly
 5000 identified linear polarization patches 8 \% were entirely
 embedded in downflows over their whole lifetime. This rules out
 emergence from below the solar surface as their origin. These
 observations together with magnetoconvection simulations prompted us
 to explore a new mechanism for transporting flux into the INW
 photosphere. This mechanism would produce naturally linear
 polarization patches entirely embedded in downflows. In the proposed
 mechanism magnetic field is dragged down from a magnetic canopy
 \citep{Gabriel1976} by downflows due to convective overshoot.

\section{MHD simulation and studying the topology}

 The magnetoconvection simulation used in this paper is the same one
 as in \citet{Pietarila+others2010}, but run for a more extended
 period of solar time. We use the MuRAM code (MPS/University of
 Chicago Radiative MHD, \citealt{Voegler+others2005}) which takes into
 account effects of full compressibility, open boundary conditions as
 well as non-gray radiative transfer and partial ionization. The
 simulation domain is a 24Mm$\times$24Mm$\times$1.68Mm
 (576$\times$576$\times$120 grid points) box of which approximately
 700 km is above the $log(\tau_{500nm})=0$ level. The boundary
 conditions are periodic in the horizontal direction. The upper
 boundary condition for the magnetic field is potential (vertical at
 infinity) and the velocity obeys an open boundary condition both at
 the top and bottom of the domain. The simulation was started from a
 snapshot of a non-magnetic convection run that had reached a statistical steady state, on to which we added a
 magnetic field. The initial configuration of the field was a strip of
 magnetic field with a squared Gaussian profile located in the center
 of the domain in the x-direction and aligned along the y-axis. We
 chose a peak value of 300 G for the strip. This way the initial
 magnetic field is energetically important above the surface but much
 less so below the surface. The full width at half maximum was chosen
 to be 3.157 Mm, so that the patch of magnetic field spans several
 granules. This setup is reminiscent of an isolated, unipolar network
 lane. The simulation was run for $\approx$ 1 hour of solar
 time. Since we are interested in the region outside the initial strip
 of magnetic field, we take advantage of the periodic boundary
 conditions in the horizontal directions and place the strip at the
 edges of the domain in the figures instead of having it in the center
 as in \citet{Pietarila+others2010} (i.e., from now on x=0 and x=24 Mm
 refer to the center of the initial strip).

\section{Results}

\begin{figure*}
\includegraphics[width=14cm]{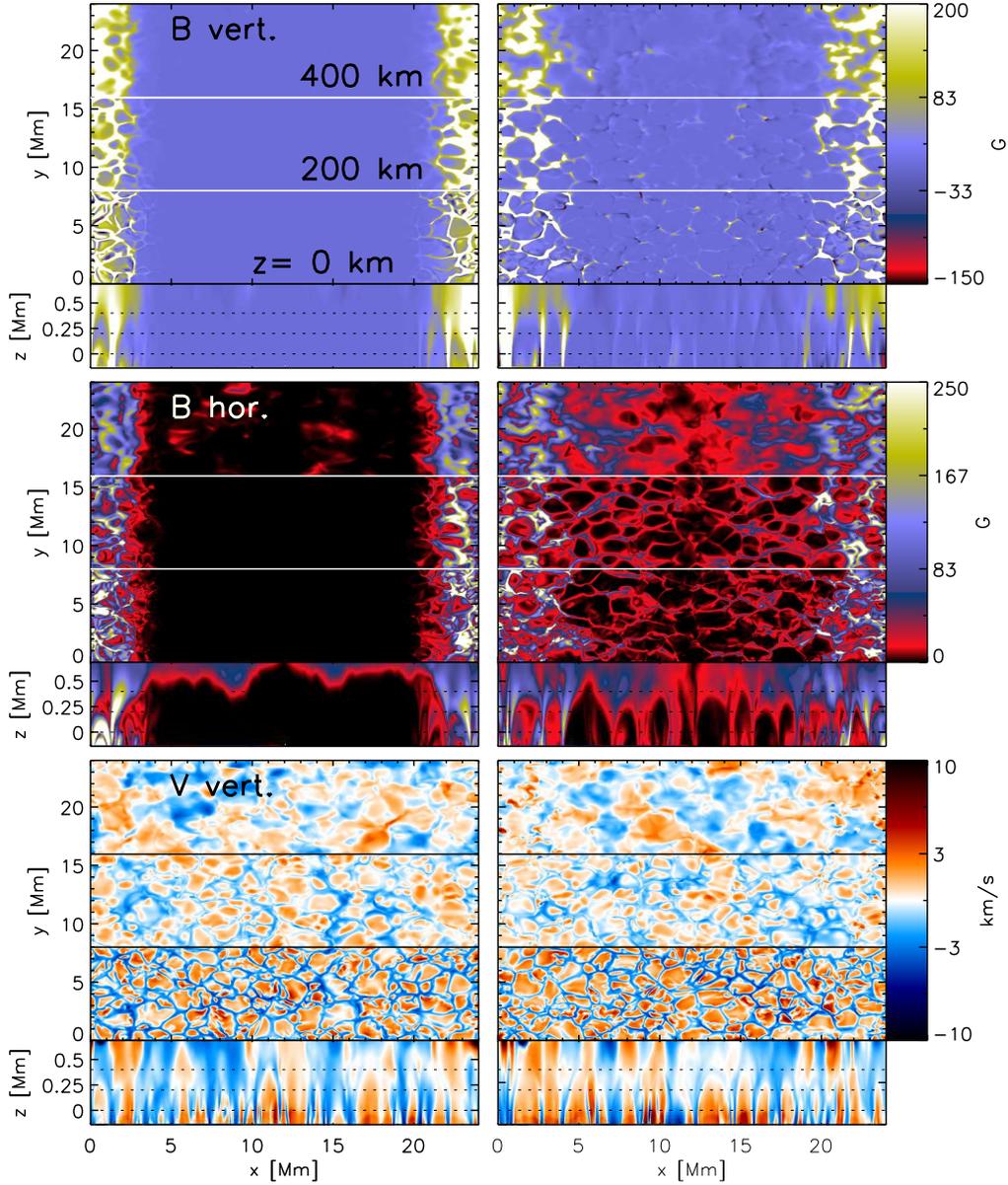}
\caption{Magnetic field and vertical velocity close to the beginning
  ($t=1.5$ min, panels on the left) and end ($t=45$ min, panels on the
  right) of the simulation. Sub-panels in top panel from top to bottom
  show x-y slices of the vertical magnetic field strength at 3
  different heights (400, 200 and 0 km) and a x-z slice of the
  simulation domain. $z=0$ km is defined as where the average
  temperature (over the entire simulation domain) at $t=0$ is
  $\approx$ 6000 K. The dashed lines in the bottom sub-panel mark the heights of the x-y
  slices. Initial distribution of field corresponds to unipolar (positive polarity) field at the left and right boundaries of the computational domain with field-free area between them. The two panels below show the same for the horizontal field
  and the vertical velocity, respectively.}
\label{fig:bigsnap}
\end{figure*}

 In the beginning of the simulation (left panels in
 Fig.~\ref{fig:bigsnap}) the vertical magnetic field is confined to
 the initial magnetic field strip. The potential field upper boundary
 condition allows the magnetic field to expand near the top of the
 domain and a canopy-like structure composed of nearly horizontal
 field forms rather quickly over the initially field-free
 regions. (From here on the term ``canopy'' is used for this
 structure.) Since the initial field is unipolar and we have periodic
 boundary conditions in the horizontal directions, the canopy field
 expands in a ``wine glass'' shape: the field is vertical at $x=12$ Mm
 where the field from two same-polarity network lanes meets. The
 spreading out of the field is apparent in the magnetic field azimuth:
 it is predominantly positive in the left ($x=0-12$ Mm) and negative
 in the right ($x=12-24$ Mm) halves of the box. The effect of
 granulation on the spatial distribution of the magnetic field is
 visible early on: the field is gathered in the intergranular downflow
 lanes. As the simulation progresses the magnetic field fills up the
 entire domain (right panels of Fig.~\ref{fig:bigsnap}). At all
 heights the granulation pattern is visible in the spatial
 distribution of the magnetic field: field is found predominantly in
 the downflow lanes. The plasma-$\beta$ is well above unity ($\ge$ 10)
 in most of the simulation domain, only close to the very top does it
 decrease to below unity. The field azimuth has the same spatial
 asymmetry (goes from positive to negative with {\it x}) at all
 heights.

\begin{figure*}
\includegraphics[width=14cm]{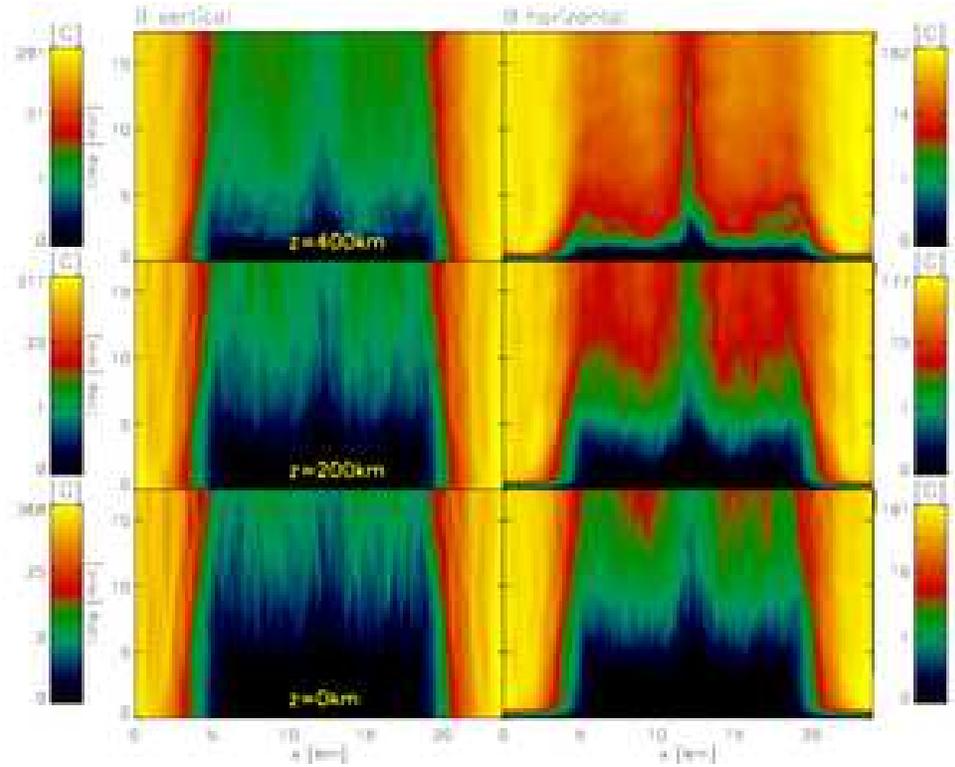}
\caption{Time evolution of the mean (averaged along y-direction)
  vertical (left) and horizontal (right) magnetic field at z=400, 200
  and 0 km during the first 18 minutes of the simulation. }
\label{fig:absb}
\end{figure*}

 The magnetic field is transported to the initially nearly field-free
 regions of the photosphere by two mechanisms, namely by horizontal
 diffusion (random walk) from the initial strip and by a process where
 the downflows due to the convective overshoot pull the field down
 from the canopy. With the simulation resolution, $\simeq$41 km in the
 horizontal direction, we do not achieve magnetic Reynolds numbers
 high enough to have local dynamo action. Therefore, we can exclude it
 as a source for the field in the initially nearly field-free
 regions. This does not exclude stretching and inductive amplification
 of the unsigned flux as a source in the later stages of the
 simulation. As we will later show when discussing the time evolution
 of the magnetic field topology this is an important process in
 supplying field to photospheric regions outside the initial strip. The prevalence
 of the spatial azimuth asymmetry throughout the simulation, both in
 height and time, points to a strong connection with the canopy which has the same asymmetry. This
 connection is also seen in the time evolution of the magnetic field
 (Fig.~\ref{fig:absb}). In the initially field-free regions, the
 magnetic field appears at a given height nearly simultaneously at all
 locations in the horizontal plane. The fine structure of the field
 has a spatial scale of about 1 Mm, corresponding to the granulation
 scale. The field fills the volume starting from the top of the domain
 and reaches the surface after $\approx 10$ minutes, i.e, after a few
 granulation timescales. The horizontal diffusion of the field from
 the initial strip is seen as the slight increase with time of the width
 of the strong field region at $x \approx 0$  and $x \approx 24$. Its
 effect in filling the volume with field at distances greater than a
 couple of Mm from the strip is negligible compared to the flux
 transported down from the canopy. The nature of the downflows and canopy-like magnetic fields will be discussed in Section ~\ref{sec:disc}.

\begin{figure*}
\includegraphics[width=14cm]{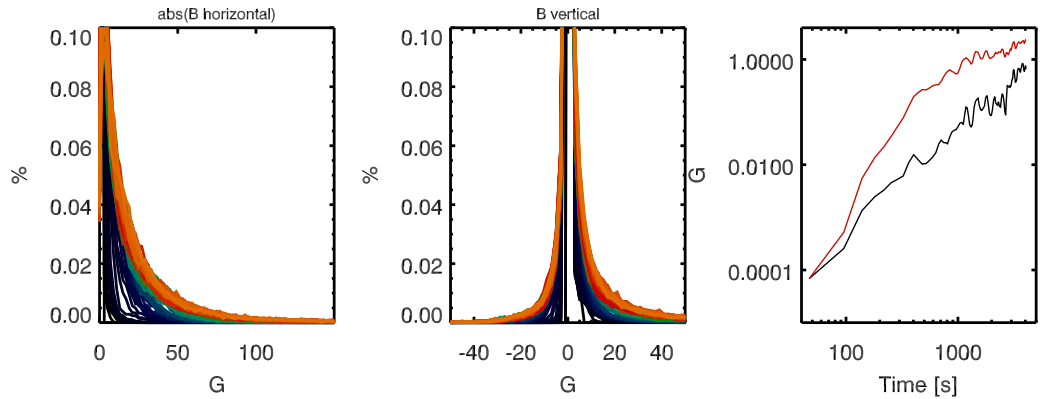}
\caption{Histograms of the horizontal (left) and vertical (middle)
  magnetic fields at $z=220$ km for each time step of the simulation
  (blacks correspond to the beginning and reds to the end of the
  simulation) in a strip at $x=4-10$ Mm of the
  simulation domain (i.e., outside the initial strip of magnetic
  field). Right plot shows the time evolution of the signed (in black)
  and unsigned (in red) vertical magnetic flux in the strip. }
\label{fig:fields}
\end{figure*}

The dominance of downflows over horizontal diffusion in providing flux close to the surface is also seen in histograms of the magnetic field
 strength outside the initial strip taken at
 different times during the simulation (two first panels in Fig.~\ref{fig:fields}). They show the strengthening of the horizontal and vertical fields. The vertical field is not unipolar as
 one would expect if the main transportation mechanism was horizontal
 diffusion. In fact, the signed vertical flux is significantly smaller
 than the unsigned flux at all times (right panel in Fig.~\ref{fig:fields}). There are indications of
 saturation of the downflow mechanism after $\approx$ 600 seconds when
 the signed flux increase becomes faster than the increase of the
 unsigned flux.

\begin{figure*}
\includegraphics[width=14cm]{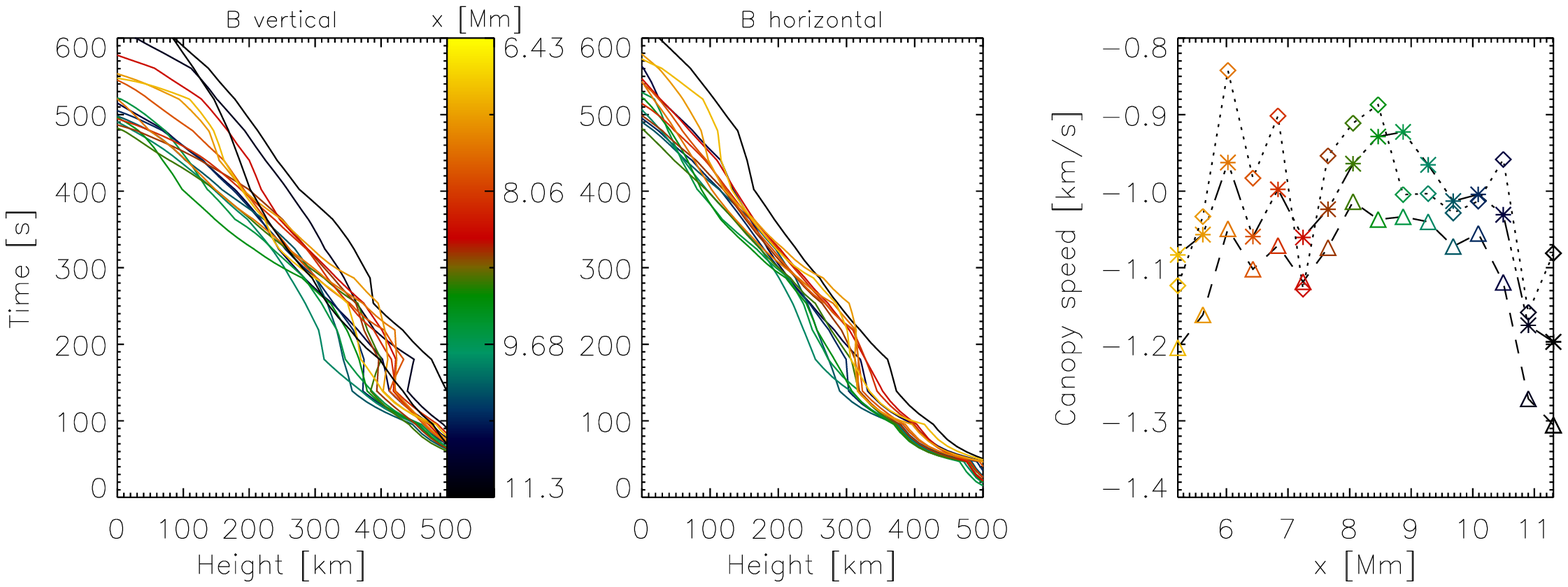}
\caption{Transport of magnetic fields by downflows. Left and middle
  panels show the time when the base of the canopy (vertical and horizontal components) reaches a given height. The
  canopy base is considered to have reached a height when at that height
  the average (over a 400 km wide strip) vertical field strength is
  above 0.33 G or  the horizontal field exceeds 0.66 G. The different
  colors are for different positions along the x-axis (shown
  in the color bar). The right panel shows the average downward velocity of the
   canopy base as a function of position along the $x$-axis. The velocity is derived from a linear fit between z=$0$ and $500$ km to the curves in the two other panels. Dotted line is for horizontal, dashed for vertical and dash-dotted for total field.  }
\label{fig:canvel}
\end{figure*}

 To quantify the apparent speed at which the horizontal magnetic field is
 advected down from the canopy we look at when the average (in
 y-direction) total field strength at a given height becomes larger
 than 1 G or the unsigned vertical field exceeds 0.33 G or the
 horizontal field 0.66 G (Fig.~\ref{fig:canvel}). The limit was chosen based
 on images similar to Fig.~\ref{fig:absb} where 1 G outlines well the
 downward moving canopy. The thus computed average downward  speed of the canopy's base is found to be $\approx$1 km
 s$^{-1}$. The simulation domain
 extends $\approx$ 700 km above $\log(\tau) \approx 0$ in height so an
 average speed of 1 km s$^{-1}$ is consistent with field from the canopy reaching
 the solar surface in $\approx$ 10 minutes.

\begin{figure*}
\includegraphics[width=14cm]{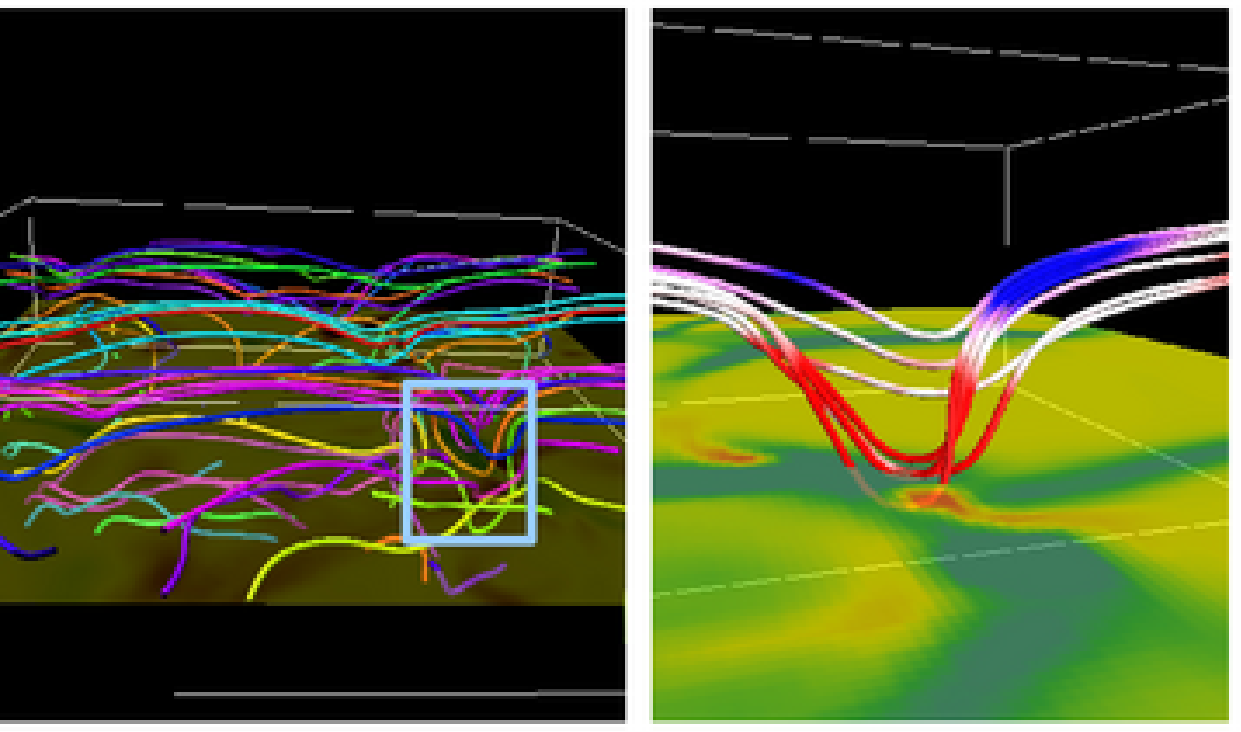}
\caption{Left: Volume rendering of magnetic field lines in a subvolume
  of the simulation domain ($x=6-10$ Mm, $y=2-6$ Mm and $z=-0.5-0.7$ Mm). The
  snapshot is at $t=14.7$ min. Right: zoom into the region outlined by the
  light blue box. The surface is the ${\rm B_{x}}$ component of the
  magnetic field at $z=0$ km. Color of the field lines in the right panel
  shows the vertical velocity (red is a downflow, blue an
  upflow. Color scale saturates at $\pm$1 km s$^{-1}$). The volume rendering was made using the Visualization and Analysis Platform for Ocean, Atmosphere, and Solar Researchers (VAPOR, www.vapor.ucar.edu)}
\label{fig:loops}
\end{figure*}

 If the magnetic field is dragged down by downflows one would expect a
 prevalence of U-loops in the field reaching the surface,
 especially in the beginning of the simulation before the field
 topology becomes more complex as convection moves the dragged-down 
  field in different directions. A 3-dimensional rendering of the
 magnetic field confirms this as is shown in Fig.~\ref{fig:loops}. Color
 coding of the field lines represents vertical velocity (with red signifying downflows). 
 It shows clearly that the U-loop is associated
 with a downflow.

 Since we are interested in the interaction of the canopy and the
 convective motions, we also performed an analysis to characterize the
 flux in the lower to mid photospheric layers as belonging to U-loops,
 $\Omega$-loops, or field lines going from below the surface to above
 the higher photospheric layers. For this purpose we define lower and
 upper surfaces at $z=-294$ and $z=406$ km. For each grid point
 $(x,y,z)$ within this domain the field through the point is traced in
 both directions until it leaves the subdomain.  We label all the
 points along the field line as U-type if both ends of the field line
 exit the subdomain through the upper plane; the points are labeled as
 $\Omega$-type if both ends of the field line exit through the bottom
 plane; the field lines whose one end of which passes through the
 bottom plane and other end through the top plane are defined as being
 either positive or negative polarity depending on the orientation of
 the field.

\begin{figure*}
\includegraphics[width=7cm]{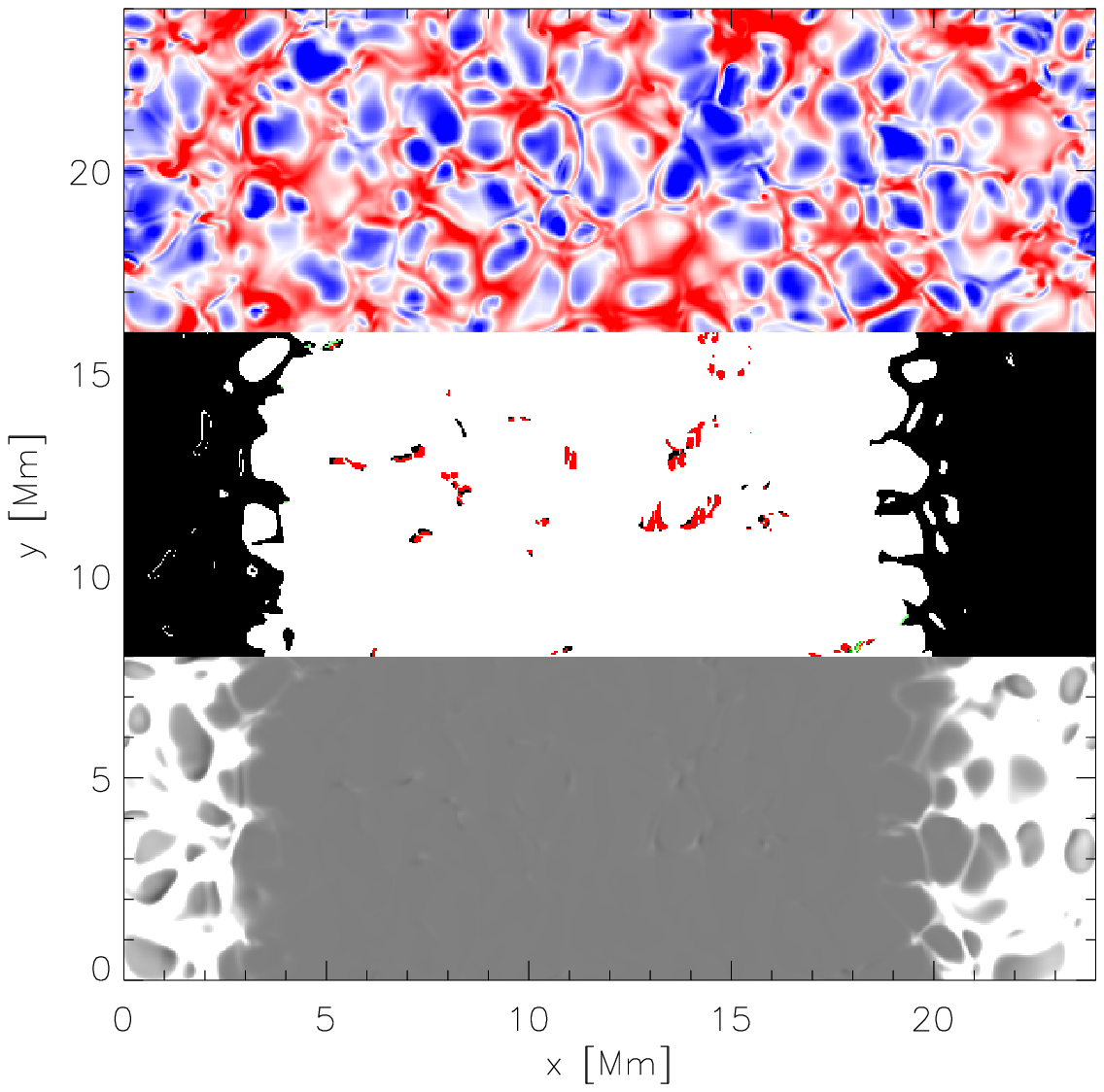}\includegraphics[width=7cm]{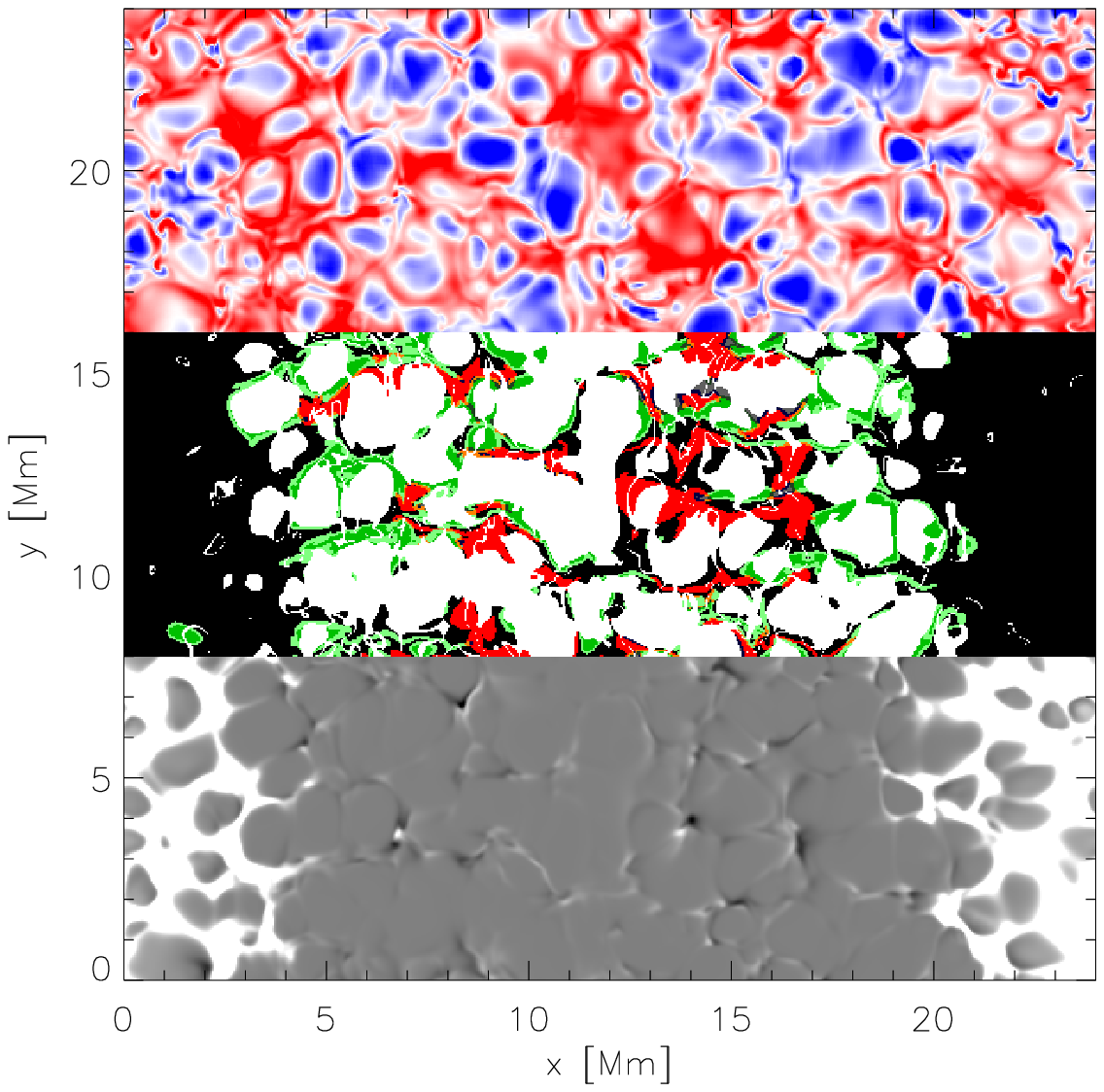}

\includegraphics[width=7cm]{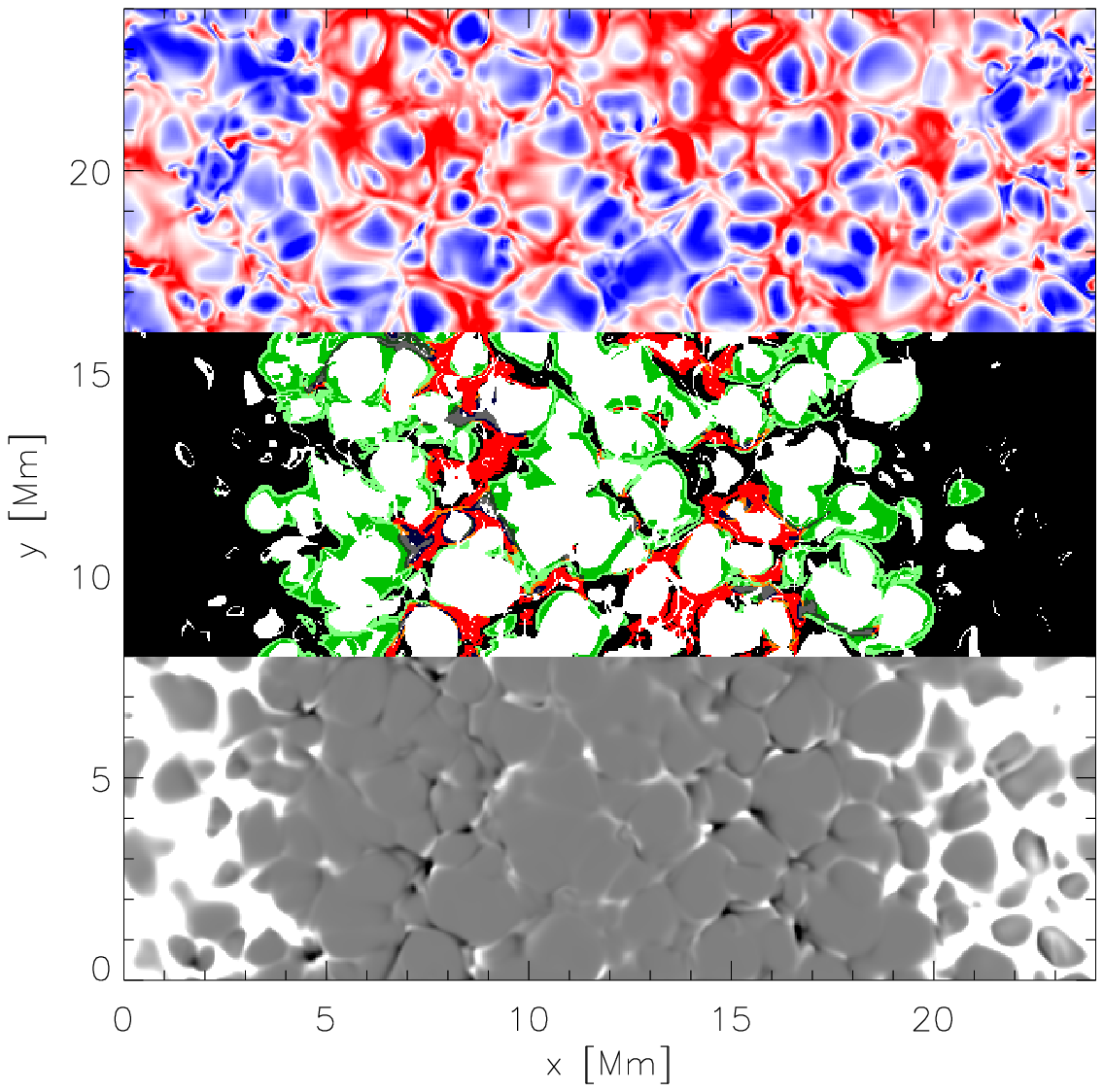}\includegraphics[width=7cm]{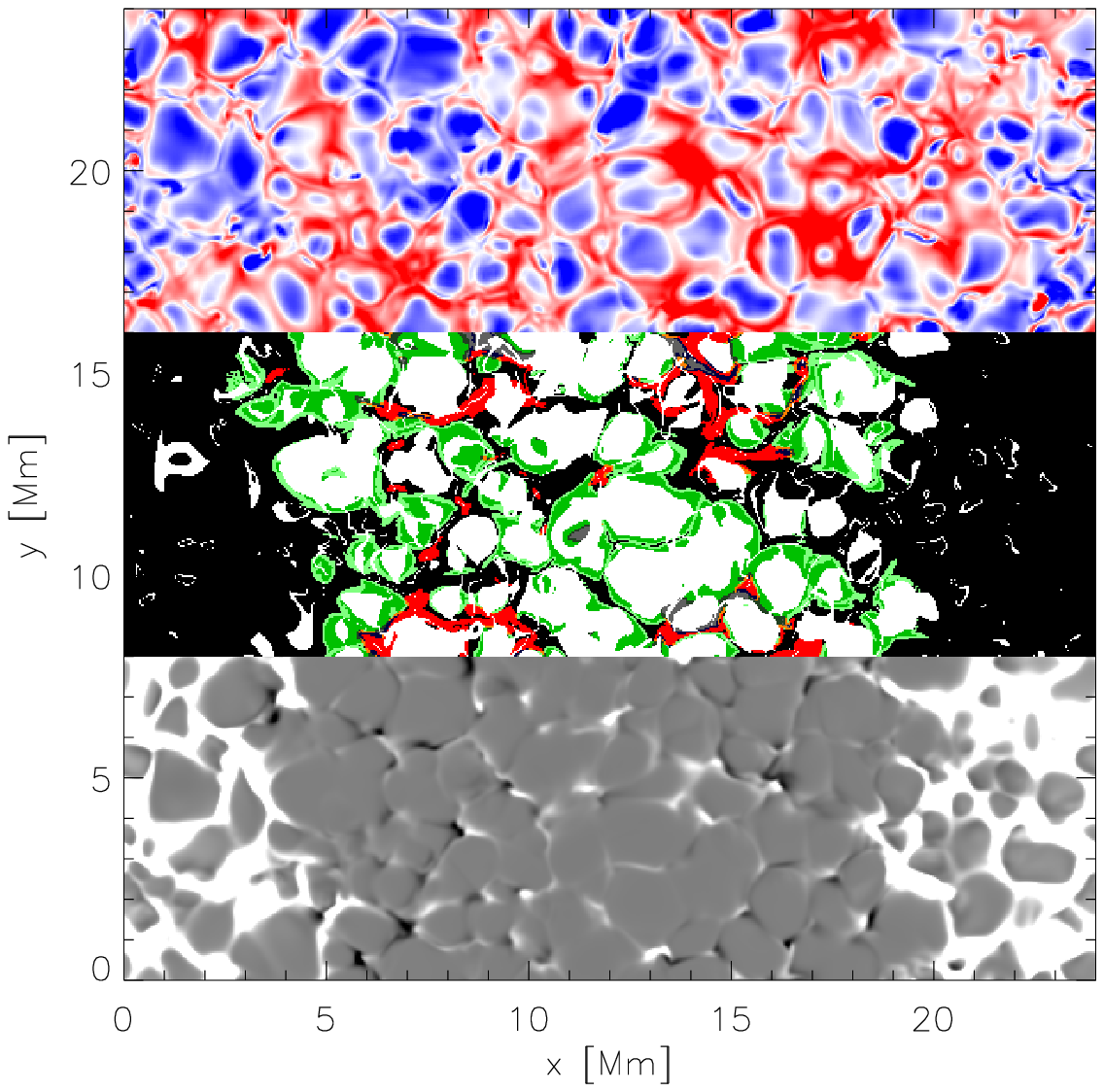}
\caption{Snapshots from the simulation at a height of 224~km at times
  t=4.2 (upper left), 14.1 (upper right), 27.3 (lower left), and 45
  minutes (lower right).  For each time the upper part of the panel
  shows the vertical component of the velocity saturated at $-2$ km
  s$^{-1}$ (blue) and $2$ km s$^{-1}$ (red).  The middle parts are
  maps of the distribution of the different types of loops.  U-type
  loops (field lines both ends of which leave through the top of the
  subdomain $-294$ km $\le z \le 406$ km are shown in red),
  $\Omega$-loops (both ends of which leave through the bottom of the
  subdomain) in green, field lines passing from the bottom to the top
  of the subdomain in black, and in the opposite direction in
  grey. The types of loops are only shown where the field exceeds
  $1$~G, all other regions are colored white.  The bottom third of
  each snapshot shows the vertical component of the magnetic field,
  saturated at $\pm$ 50 G.  }
\label{fig:loopcount}
\end{figure*}

The resulting classification of the magnetic field at z=$224$~km is
shown for four timesteps in Fig.~\ref{fig:loopcount}. The first
snapshot is at 4.2 minutes, i.e., when the downflowing canopy
reaches this height for the first time. At this time the downflows
have had sufficient time to pull parts of the horizontal canopy down
into the lower photosphere. The INW flux is located in the junctions
of the intergranular lanes and it is nearly entirely in the form of
U-loops or positive polarity field passing through the domain. The
positive polarity field is due to flux expulsion: as 4.2 minutes is
comparable to the lifetime of the granulation there has been enough
time for flux expulsion to concentrate the very weak field in the
region from $x=5$ to $20$ Mm into a number of isolated elements which
thread the subdomain from bottom to top (and hence appear black in the
topology maps). At later times, 14.1, 27.3 min, and 45 min, there are
still U-loops present in the INW, but also a large amount of
$\Omega$-loops are seen as well as positive polarity field.  The
U-loops appear as small bipoles in the bottom sub-panels showing the
vertical field, with a vertical magnetic field strength which is
generally stronger than in the $\Omega$-loops and in the vertical INW
field. Because we do not have a supergranular scale flow in the
simulation, flux which was originally in the unipolar network features
begins to get concentrated into granular downflow lanes while at the
same time gradually diffusing outward from its original location, as
can be seen in the snapshot from 45 minutes.

\begin{figure*}
\includegraphics[width=7cm]{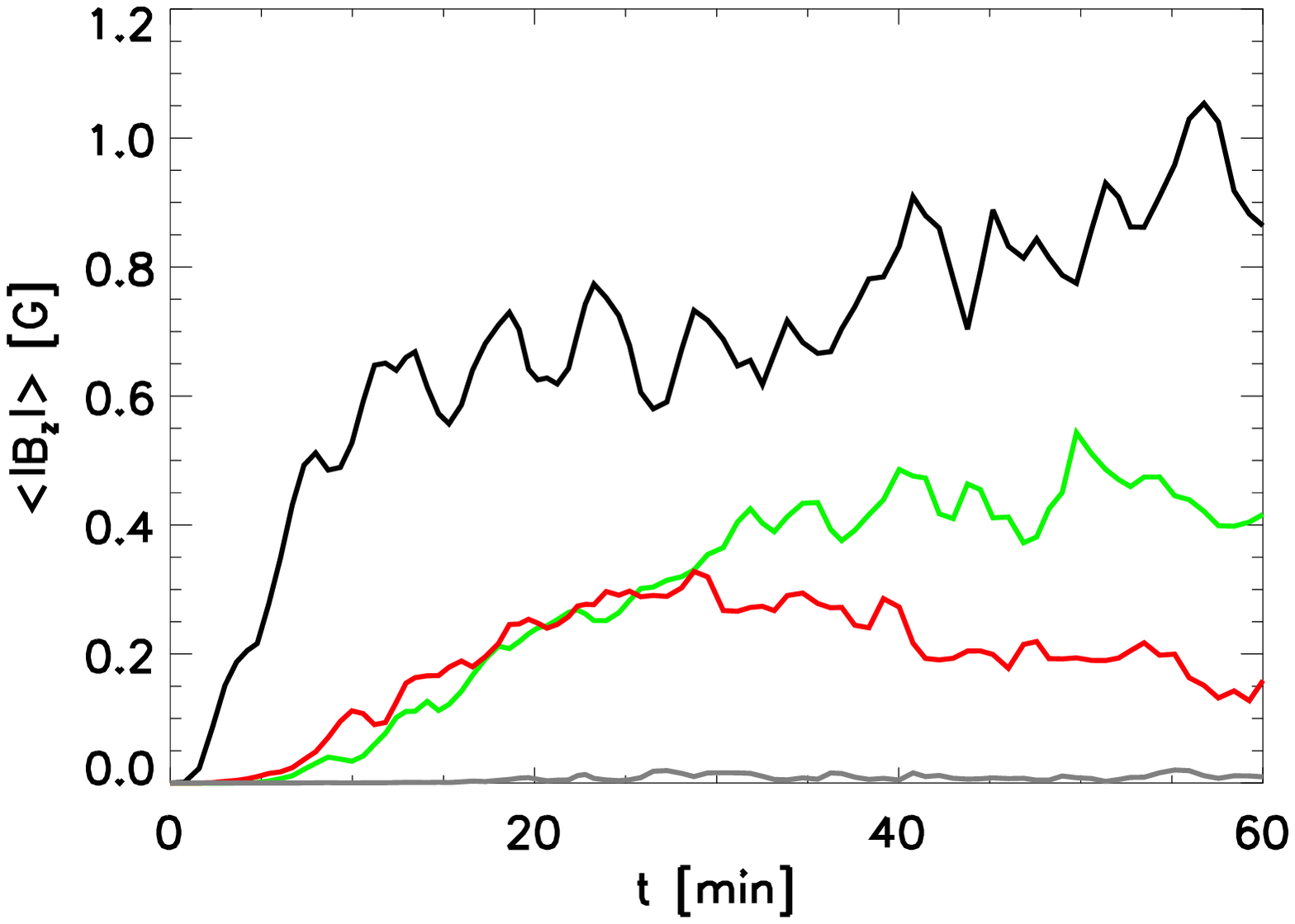}\includegraphics[width=7cm]{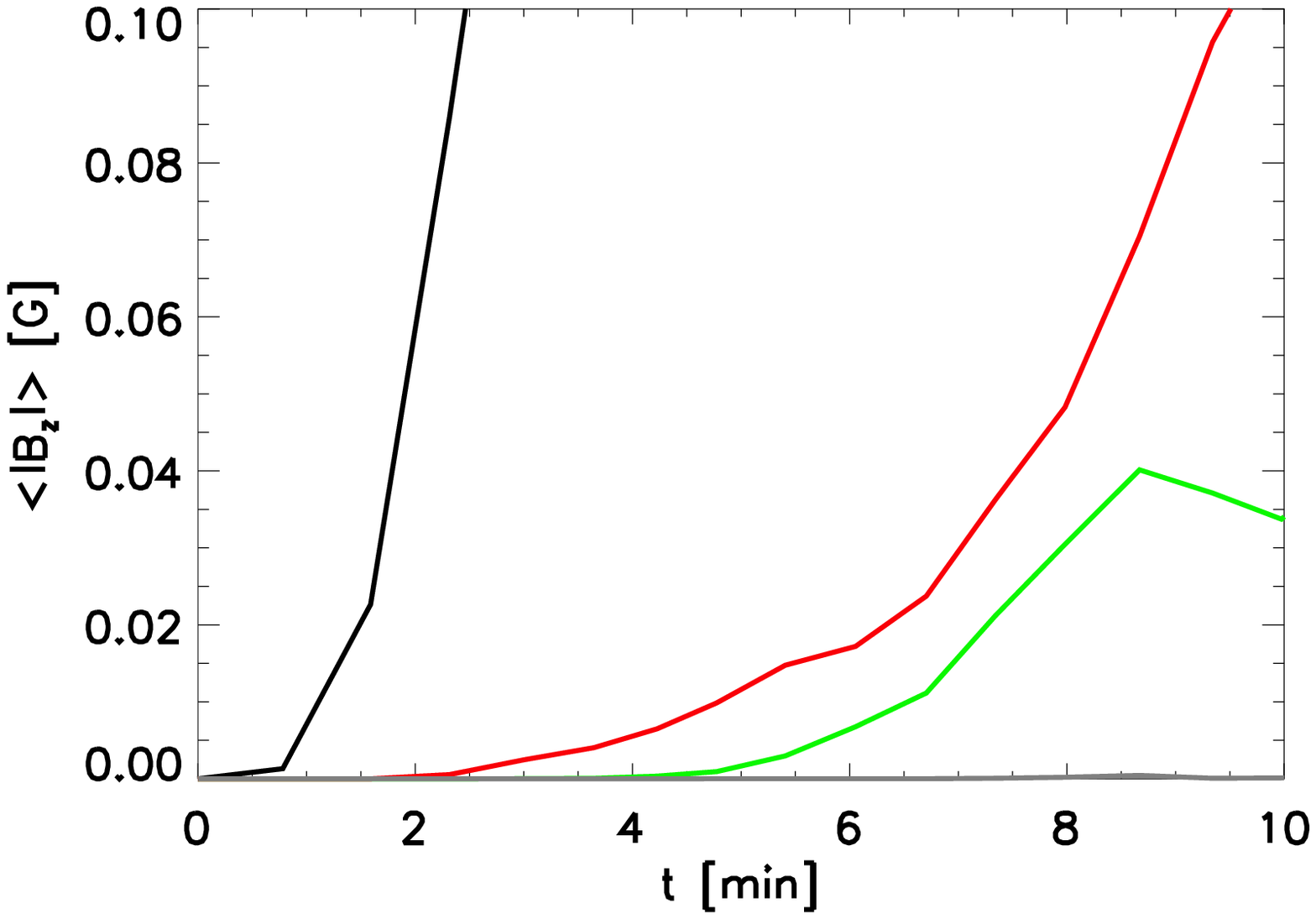}

\caption{ The time evolution of the different types of spatially averaged 
  flux densities at height
  224~km. The classification into U, $\Omega$ or straight-through
  topology is based on the subdomain $-294 \le z\le 406$~km.  The
  amount of \black unsigned \black flux associated with U-type loops is shown in red,
  $\Omega$-loops in green, field lines passing from the bottom to the
  top of the subdomain in black, and in the opposite direction in
  grey.  The total amount of signed flux passing through the box has
  been subtracted from the black curve, \black so that the curve gives the increase in
      unsigned flux due to turbulent stretching and tangling of the
      field within the subdomain\black. The right panel shows a zoom
  into the lower left corner of the left panel.  }
\label{fig:time_evol}
\end{figure*}

Figure~\ref{fig:time_evol} displays the time evolution of the average
unsigned flux density in the loops and single polarity field.
{\black{Throughout the simulation most of the increase in unsigned flux
    in the volume within the height range $-294$~km~$\le z \le 406$~km
    is from field lines which enter the volume from $z=-294$~km and
    leave through $z=406$~km. In between these two heights the
    unsigned field lines may be rather tangled and form small loops
    which contribute to the net unsigned flux. Since the field lines
    do not again leave the subdomain through the bottom they are not
    classified as loops. Unsigned flux of this type can increase
    through interaction with the near-surface turbulent eddies, and
    represents the pulling of the lowest lying flux above the surface
    into the convection zone. Higher lying flux is then pulled down,
    forming U-loops both ends of which pass through the plane
    $z=406$~km.} \black} During the first 10 minutes of the simulation
    U-loops dominate over $\Omega$-loops. 
They remain an important factor
also in the later stages of the simulation. 
As the field in the originally non-magnetic region builds up, field lines
with different topologies can interact and reconnect. The reconnection occurs
preferentially in the higher levels where the lower pressures allow the
flux tubes to expand horizontally and, therefore, more easily come into contact
with other flux tubes. For this reason the reconnection tends to decrease the depth
to which U-loops penetrate. Furthermore, magnetic tension 
can more easily pull the shallow, reconnected, U-loops out of the simulation domain. This may explain the decrease 
of flux associated with U-loops in Figure~\ref{fig:time_evol}. Small-scale inductive
stretching of the INW field lines over several turn-over times of the
granulation {\black can pull the now tangled field lines below the
  $z=-294$~km plane explaining \black} the appearance of the
$\Omega$-loops (green in Figs.~\ref{fig:loopcount} and
~\ref{fig:time_evol}).  {\black The average unsigned flux densities shown
  in Figure~\ref{fig:time_evol} are expected to depend strongly on the
  amount of signed flux in the network lane. The values are also
  likely affected by the choice of a potential field boundary
  condition and the height at which this boundary is placed.
  Furthermore, the amount of tangling within the photosphere in the
  simulations depends on the magnetic Reynolds number, with
  small-scale dynamo action occurring for sufficiently high values
  \citep{Vogler+Schussler2007,PietarilaGraham+others2010}. For these
  reasons, while it is qualitatively clear that the canopy can supply
  a significant amount of flux to the photosphere, it is difficult to
  estimate the amount quantitatively from the present simulations. \black}

\begin{figure*}
\includegraphics[width=14cm]{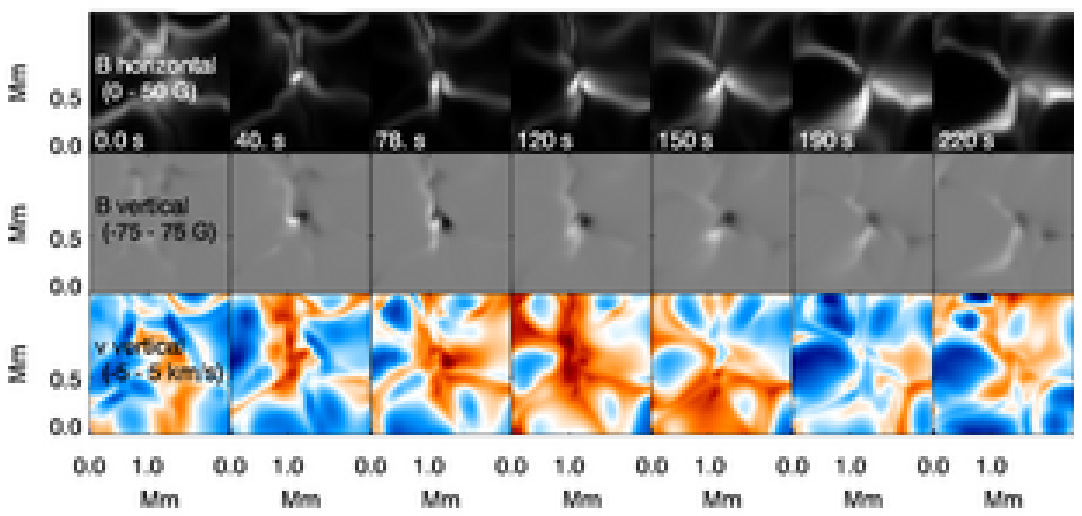}
\caption{Time evolution of magnetic field and velocity at $z=220$ km in
  a region with magnetic flux appearing in the form of a U-loop
  from above. Top row: Horizontal magnetic field strength, middle row:
  vertical magnetic field strength, bottom row: vertical
  velocity (red corresponds to downflows and blue to upflows). Times for snapshots are given in the bottom of the top row
  images. The region is centered at $x=13.5$ Mm and $y=3.7$ Mm. }
\label{fig:snaps}
\end{figure*}

 Figure ~\ref{fig:snaps} shows the time evolution of the horizontal
 and vertical magnetic fields and the vertical velocity of the U-loops
 shown in Fig.~\ref{fig:loops}. In the beginning there is only a
 little magnetic field in the junction of the intergranular lanes, but
 as the downflow plume reaches this height the field becomes
 stronger. The area with magnetic field becomes larger with time,
 consistent with the U-loop becoming more submerged. The U-loop is
 stretched and distorted as the field is further dragged into the
 downflow lanes. This is clearly seen in the fanning out of the
 horizontal field along the downflow lanes. After some time as the
 granulation evolves the field occupies a large fraction of the
 downflow lanes and is no longer as strong or as localized. The
 horizontal field strength in the feature is over 80 G at its
 strongest. The vertical field strengths are slightly lower, around 60
 G. The size of the feature before it is spread out and weakened is
 less than an arcsecond across. This is a typical example of a submerging U-loop.

\section{Discussion}
\label{sec:disc}
 In this paper we have looked at a scenario for transporting magnetic
 flux to the quiet Sun INW surface: already existing magnetic field is
 dragged down from a canopy-like configuration (consisting of mostly
 horizontal magnetic field) by downflows related to convective
 overshoot. An observational signature of the mechanism would be
 patches of linear polarization entirely embedded in downflows. Such
 patches were found to be common (8\% of all linear polarization
 patches identified in the study) in the quiet-Sun INW by
 \cite{Danilovic+others2010}.

\nocite{Abbett2007}

A similar process has been reported by Abbett (2007, in particular
Fig.10 and the text describing it). The current study focuses on the canopy fields from
network elements whereas Abbett studied convective dynamo action in the
presence of an initially uniform field. While the rates and
importance of the processes will differ in the two cases, the occurrence
of the process in the two simulations suggests that it is
robust. Importantly we note that Abbett (2007) simulations extended to
coronal heights with a simpler treatment of the radiative losses. It
should also be noted that the proposed mechanism is closely related to
that of downward pumping as described by
\cite{Thomas+others2002,Brummell+others2008}, however, now applied to
the INW surrounded by network magnetic field and overlaid by a canopy.

In the simulation nearly all of the horizontal field is in a high plasma-$\beta$ regime. If the field was stronger, it would appear more rigid to the downflows and, thus, would also less likely be dragged down. The horizontal field near the top of the
simulation domain (our canopy) should be thought of as weaker flux lying beneath the actual chromospheric canopy as inferred from observations. This weaker canopy is there because the chromospheric canopy's lower boundary is not sharp
but is rather disturbed by shocks, overshooting convection, propagating waves and turbulence.

Several aspects of our simulation are idealized, especially the potential field boundary condition imposed at a relatively
low height. The mechanism we have investigated, however, is likely to take place in the real Sun as
well if there is field associated with the overlying canopy at heights reached by the overshooting convection. The field dragged down below the surface by the downward plumes could then act as an additional source for the
generation of more field.

\cite{Danilovic+others2010} present statistics of linear polarization
patches in an INW region observed with the IMaX instrument on
Sunrise. They find that 8\% of the observed linear polarization
patches are associated purely with signatures of
submergence. Moreover, 16\% of all the observed patches begin in a
downflow. These features are not consistent with emergence of flux
from below, where an upflow would initially be expected. However, they
are compatible with the scenario proposed here. Furthermore, the sizes
and field strengths of the observed linear polarization patches are
comparable with the features seen in the simulation. Of course, we
cannot rule out other possible explanations for the linear
polarization features embedded in downflows.\black For example, the
local dynamo simulations suggest the existence of very small-scale
loops with footpoints embedded in intergranular lanes
\citep{Vogler+Schussler2007,Schussler+Vogler2008}. Their polarimetric
signature can vary as they are redistributed by the turbulent flows
\citep{Danilovic+others2010b}. In addition, $\Omega$-loops, emerged or
generated by reconnection above the solar surface, can be pulled back
down by convective motions \citep{Chae+others2004,Stein+Nordlund2006} producing the
linear polarization features associated with downflows as a result.\black

It is difficult for a number of reasons to conclusively identify the mechanism proposed here in observations of the real Sun. We expect a
 mixture of processes to transport magnetic field to the INW and only a small portion of the observed linear
 polarization patches are embedded in downflows. In the simulation the cleanest examples are seen in the
 beginning when there is only little flux outside the initial strip. As the simulation evolves, the magnetic
 field configuration becomes more complex. One possible observational
 test would be to look for spatial patterns in the azimuth of INW
 linear polarization patches embedded in downflows. This
 would, however, require solving the 180 degree ambiguity correctly.

\acknowledgements{This work was partly supported by the WCU grant
  No. R31-10016 from the Korean Ministry of Education, Science and
  Technology.}

\bibliographystyle{aa} 
\bibliography{spicules,general,fluxtube,inw,anna_rhc}

\begin{thebibliography}{29}
\expandafter\ifx\csname natexlab\endcsname\relax\def\natexlab#1{#1}\fi

\bibitem[{Abbett}(2007){Abbett}]{Abbett2007}
{Abbett}, W. P. 2007, \apj, 665, 1469

\bibitem[{{Barthol} {et~al.}(2011){Barthol}, {Gandorfer}, {Solanki}, \&
  et~al.}]{Barthol2010}
{Barthol}, P., {Gandorfer}, A., {Solanki}, S.~K., \& et~al. 2011, \solphys, 268, 1

\bibitem[{{Berkefeld} {et~al.}(2011){Berkefeld}, {Schmidt}, {Soltau}, {Bell},
  {Doerr}, {Feger}, {Friedlein}, {Gerber}, {Heidecke}, {Kentischer},
  {L{\"u}he}, {Sigwarth}, {W{\"a}lde}, {Barthol}, {Deutsch}, {Gandorfer},
  {Germerott}, {Grauf}, {Meller}, {Alvarez-Herrero}, {Kn{\"o}lker}, {Martinez
  Pillet}, {Solanki}, \& {Title}}]{Berkefeld+others2010}
{Berkefeld}, T., {Schmidt}, W., {Soltau}, D., {et~al.} 2011, \solphys, 268, 103

\bibitem[{{Brummell} {et~al.}(2008){Brummell}, {Tobias}, {Thomas}, \&
  {Weiss}}]{Brummell+others2008}
{Brummell}, N.~H., {Tobias}, S.~M., {Thomas}, J.~H., \& {Weiss}, N.~O. 2008,
  \apj, 686, 1454

\bibitem[{{Centeno} {et~al.}(2007){Centeno}, {Socas-Navarro}, \&
  {Lites}}]{Centeno+others2007}
{Centeno}, R., {Socas-Navarro}, H., \& {Lites}, B. 2007, \apjl, 666, L137


\bibitem[{Chae} {et~al.}(2004){Chae}, {Moon}, \& {Pevtsov}]{Chae+others2004}
{Chae}, J., {Moon}, Y.~J., \& {Pevtsov}, A. A. 2004, \apjl, 602, L65


\bibitem[{{Danilovic} {et~al.}(2010{\natexlab{a}}){Danilovic}, {Beeck},
  {Pietarila}, {Sch{\"u}ssler}, {Solanki}, {Mart{\'{\i}}nez Pillet}, {Bonet},
  {del Toro Iniesta}, {Domingo}, {Barthol}, {Berkefeld}, {Gandorfer},
  {Kn{\"o}lker}, {Schmidt}, \& {Title}}]{Danilovic+others2010}
{Danilovic}, S., {Beeck}, B., {Pietarila}, A., {et~al.} 2010{\natexlab{a}},
  \apjl, 723, L149

\bibitem[{{Danilovic} {et~al.}(2010{\natexlab{a}}){Danilovic}, {Sch{\"u}ssler},
  \& {Solanki}}]{Danilovic+others2010b}
{Danilovic}, S., {Sch{\"u}ssler}, M., \& {Solanki}, S.~K. 2010{\natexlab{b}},
  \aap, 513, A76

\bibitem[{{de Wijn} {et~al.}(2009){de Wijn}, {McIntosh}, \& {De
  Pontieu}}]{deWijn+others2009}
{de Wijn}, A.~G., {McIntosh}, S.~W., \& {De Pontieu}, B. 2009, \apjl, 702, L168

\bibitem[{{Gabriel}(1976){Gabriel}}]{Gabriel1976}
{Gabriel}, A.~H. 1976, Royal Society of London Philosophical Transactions Series A, 281, 339

\bibitem[{{Gandorfer} {et~al.}(2011){Gandorfer}, {Grauf}, {Barthol},
  {Riethmueller}, {Solanki}, {Chares}, {Deutsch}, {Ebert}, {Feller},
  {Germerott}, {Heerlein}, {Heinrichs}, {Hirche}, {Hirzberger}, {Kolleck},
  {Meller}, {Mueller}, {Schaefer}, {Tomasch}, {Knoelker}, {Martinez Pillet},
  {Bonet}, {Schmidt}, {Berkefeld}, {Feger}, {Heidecke}, {Soltau},
  {Tischenberg}, {Fischer}, {Title}, {Anwand}, \&
  {Schmidt}}]{Gandorfer+others2010}
{Gandorfer}, A., {Grauf}, B., {Barthol}, P., {et~al.} 2010, \solphys, 268, 35

\bibitem[{{Hagenaar} {et~al.}(2003){Hagenaar}, {Schrijver}, \&
  {Title}}]{Hagenaar+others2003}
{Hagenaar}, H.~J., {Schrijver}, C.~J., \& {Title}, A.~M. 2003, \apj, 584, 1107

\bibitem[{{Kosugi} {et~al.}(2007){Kosugi}, {Matsuzaki}, {Sakao}, \&
  et~al.}]{Hinode}
{Kosugi}, T., {Matsuzaki}, K., {Sakao}, T., \& et~al. 2007, \solphys, 243, 3

\bibitem[{{Lites} {et~al.}(2001){Lites}, {Elmore}, \& {Streander}}]{HinSP2001}
{Lites}, B.~W., {Elmore}, D.~F., \& {Streander}, K.~V. 2001, in Astronomical
  Society of the Pacific Conference Series, Vol. 236, Advanced Solar
  Polarimetry -- Theory, Observation, and Instrumentation, ed. {M.~Sigwarth},
  33

\bibitem[{{Mart{\'{\i}}nez Gonz{\'a}lez} \& {Bellot
  Rubio}(2009)}]{MartinezGonzalez+BellotRubio2009}
{Mart{\'{\i}}nez Gonz{\'a}lez}, M.~J. \& {Bellot Rubio}, L.~R. 2009, \apj, 700,
  1391

\bibitem[{{Mart{\'{\i}}nez Gonz{\'a}lez} {et~al.}(2007){Mart{\'{\i}}nez
  Gonz{\'a}lez}, {Collados}, {Ruiz Cobo}, \&
  {Solanki}}]{MartinezGonzalez+others2007}
{Mart{\'{\i}}nez Gonz{\'a}lez}, M.~J., {Collados}, M., {Ruiz Cobo}, B., \&
  {Solanki}, S.~K. 2007, \aap, 469, L39

\bibitem[{{Mart\'inez Pillet} {et~al.}(2011){Mart\'inez Pillet}, {del Toro
  Iniesta}, {\'Alvarez-Herrero}, \& et~al.}]{MartinezPillet+others2010}
{Mart\'inez Pillet}, V., {del Toro Iniesta}, J.~C., {\'Alvarez-Herrero}, A., \&
  et~al. 2011, \solphys, 268, 57

\bibitem[{{Petrovay} \& {Szakaly}(1993)}]{Petrovay+Szakaly1993}
{Petrovay}, K. \& {Szakaly}, G. 1993, \aap, 274, 543

\bibitem[{{Pietarila} {et~al.}(2010){Pietarila}, {Cameron}, \&
  {Solanki}}]{Pietarila+others2010}
{Pietarila}, A., {Cameron}, R., \& {Solanki}, S.~K. 2010, \aap, 518, A50

\bibitem[{{Pietarila Graham} {et~al.}(2010){Pietarila Graham}, {Cameron}, \&
  {Sch{\"u}ssler}}]{PietarilaGraham+others2010}
{Pietarila Graham}, J., {Cameron}, R., \& {Sch{\"u}ssler}, M. 2010, \apj, 714,
  1606

\bibitem[{{Ploner} {et~al.}(2001){Ploner}, {Sch{\"u}ssler}, {Solanki}, \&
  {Gadun}}]{Ploner+others2001}
{Ploner}, S.~R.~O., {Sch{\"u}ssler}, M., {Solanki}, S.~K., \& {Gadun}, A.~S.
  2001, in Astronomical Society of the Pacific Conference Series, Vol. 236,
  Advanced Solar Polarimetry -- Theory, Observation, and Instrumentation, ed.
  {M.~Sigwarth}, 363

\bibitem[{{Sch{\"u}ssler} \& {V{\"o}gler}(2008)}]{Schussler+Vogler2008}
{Sch{\"u}ssler}, M. \& {V{\"o}gler}, A. 2008, \aap, 481, L5

\bibitem[{{Shimizu} {et~al.}(2008){Shimizu}, {Nagata}, {Tsuneta}, {Tarbell},
  {Edwards}, {Shine}, {Hoffmann}, {Thomas}, {Sour}, {Rehse}, {Ito},
  {Kashiwagi}, {Tabata}, {Kodeki}, {Nagase}, {Matsuzaki}, {Kobayashi},
  {Ichimoto}, \& {Suematsu}}]{Shimizu+others2008}
{Shimizu}, T., {Nagata}, S., {Tsuneta}, S., {et~al.} 2008, \solphys, 249, 221

\bibitem[{{Solanki} {et~al.}(2010){Solanki}, {Barthol}, {Danilovic}, {Feller},
  {Gandorfer}, {Hirzberger}, {Riethm{\"u}ller}, {Sch{\"u}ssler}, {Bonet},
  {Mart{\'{\i}}nez Pillet}, {del Toro Iniesta}, {Domingo}, {Palacios},
  {Kn{\"o}lker}, {Bello Gonz{\'a}lez}, {Berkefeld}, {Franz}, {Schmidt}, \&
  {Title}}]{Solanki+others2010}
{Solanki}, S.~K., {Barthol}, P., {Danilovic}, S., {et~al.} 2010, \apjl, 723,
  L127

\bibitem[{{Stein} \& {Nordlund}(2006)}]{Stein+Nordlund2006}
{Stein}, R.~F. \& {Nordlund}, {\AA}. 2006, \apj, 642, 1246

\bibitem[{{Suematsu} {et~al.}(2008){Suematsu}, {Tsuneta}, {Ichimoto},
  {Shimizu}, {Otsubo}, {Katsukawa}, {Nakagiri}, {Noguchi}, {Tamura}, {Kato},
  {Hara}, {Kubo}, {Mikami}, {Saito}, {Matsushita}, {Kawaguchi}, {Nakaoji},
  {Nagae}, {Shimada}, {Takeyama}, \& {Yamamuro}}]{Suematsu+others2008}
{Suematsu}, Y., {Tsuneta}, S., {Ichimoto}, K., {et~al.} 2008, \solphys, 249,
  197

\bibitem[{{Thomas} {et~al.}(2002){Thomas}, {Weiss}, {Tobias}, \&
  {Brummell}}]{Thomas+others2002}
{Thomas}, J.~H., {Weiss}, N.~O., {Tobias}, S.~M., \& {Brummell}, N.~H. 2002,
  \nat, 420, 390

\bibitem[{{Trujillo Bueno} {et~al.}(2004){Trujillo Bueno}, {Shchukina}, \&
  {Asensio Ramos}}]{TrujilloBueno+others2004}
{Trujillo Bueno}, J., {Shchukina}, N., \& {Asensio Ramos}, A. 2004, \nat, 430,
  326

\bibitem[{{Tsuneta} {et~al.}(2008){Tsuneta}, {Ichimoto}, {Katsukawa}, {Nagata},
  {Otsubo}, {Shimizu}, {Suematsu}, {Nakagiri}, {Noguchi}, {Tarbell}, {Title},
  {Shine}, {Rosenberg}, {Hoffmann}, {Jurcevich}, {Kushner}, {Levay}, {Lites},
  {Elmore}, {Matsushita}, {Kawaguchi}, {Saito}, {Mikami}, {Hill}, \&
  {Owens}}]{Tsuneta+others2008}
{Tsuneta}, S., {Ichimoto}, K., {Katsukawa}, Y., {et~al.} 2008, \solphys, 249,
  167

\bibitem[{{V{\"o}gler} \& {Sch{\"u}ssler}(2007)}]{Vogler+Schussler2007}
{V{\"o}gler}, A. \& {Sch{\"u}ssler}, M. 2007, \aap, 465, L43

\bibitem[{{V{\"o}gler} {et~al.}(2005){V{\"o}gler}, {Shelyag}, {Sch{\"u}ssler},
  {Cattaneo}, {Emonet}, \& {Linde}}]{Voegler+others2005}
{V{\"o}gler}, A., {Shelyag}, S., {Sch{\"u}ssler}, M., {et~al.} 2005, \aap, 429,
  335

\bibitem[{{Wedemeyer-B{\"o}hm} {et~al.}(2009){Wedemeyer-B{\"o}hm}, {Lagg}, \&
  {Nordlund}}]{WedemeyerBohm+others2009}
{Wedemeyer-B{\"o}hm}, S., {Lagg}, A., \& {Nordlund}, {\AA}. 2009, Space Science
  Reviews, 144, 317

\end{thebibliography}
\end{document}